# Intrinsic response time of graphene photodetectors

Alexander Urich[*], Karl Unterrainer, and Thomas Mueller[*]

Vienna University of Technology, Institute of Photonics,

Gußhausstraße 27-29, 1040 Vienna, Austria

[*]e-mail: alexander.urich@tuwien.ac.at, thomas.mueller@tuwien.ac.at

**Graphene-based photodetectors are promising new devices for high-speed optoelectronic applications. However, despite recent efforts, it is not clear what determines the ultimate speed limit of these devices. Here, we present measurements of the intrinsic response time of metal-graphene-metal photodetectors with monolayer graphene using an optical correlation technique with ultrashort laser pulses. We obtain a response time of 2.1 ps that is mainly given by the short lifetime of the photogenerated carriers. This time translates into a bandwidth of ~262 GHz. Moreover, we investigate the dependence of the response time on gate voltage and illumination laser power.**

Graphene, a single layer of carbon atoms arranged in a honeycomb lattice, has recently attracted enormous attention and generated intense research activity[1]. Besides its electronic properties, such as high mobility of electrons and holes or ballistic transport[2], the optical properties of graphene are currently an area of strong interest[3]. The potential of graphene in photonics and optoelectronics has been demonstrated by the realization of ultrafast lasers[4], LCD screens[5,6], solar cells[7], OLEDs[8], photodetectors[9] and other applications[3]. Graphene-based optoelectronic devices not only operate in a very wide wavelength range, but also show fast carrier transport and exhibit a short lifetime of the photogenerated carriers, resulting in a short intrinsic response time of the devices. This allows operation at very high frequencies, a feature that is particularly desirable for applications in high-speed optical communications. Operation of a metal-graphene-metal (MGM) photodetector up to a frequency of 40 GHz has recently been demonstrated[10]. In this experiment, the device worked without any performance degradation up to the frequency limit of the used measurement equipment. A limitation of the detector could hence not be determined. In this letter, we report the measurement of the intrinsic response time of MGM photodetectors with monolayer graphene, using an ultrafast optical correlation technique[11]. In contrast to a fully electronic approach, high-speed electronic equipment is not required. The experimental technique is based on the detection of the autocorrelation of ultrashort laser pulses with an MGM photodetector operated in a nonlinear regime.

The MGM photodetectors used in this experiment were fabricated from graphene that was mechanically exfoliated from natural graphite and deposited onto a highly resistive Si wafer with a layer of 300 nm $SiO_2$. Graphene monolayers were preselected with an optical microscope and subsequently characterized by Raman spectroscopy[12]. Source and drain contacts, consisting of 20 nm Ti and 40 nm Au, were defined with optical lithography and fabricated by vapor deposition. The Si substrate was electrically contacted and served as a back gate electrode. An image of a typical device is incorporated in Figure 1(a).

Figure 1(a) shows a schematic illustration of the experimental setup. A beam of ultrashort laser pulses from an erbium fiber laser with a pulse length[13] of about 100 fs, a center wavelength of 1.55 µm, and a repetition rate of 80 MHz is split into two parts and subsequently recombined. The time delay between the pulses of the beams is adjusted with a translation stage in one of the beam paths. Both beams are set to have the same pulse energy (~8 pJ). After recombination, the beams are focused with an objective lens (NA = 0.55) onto an MGM photodetector resulting in a laser spot of ~3 µm in diameter. The photodetector is mounted on an X-Y translation stage with a resolution of 100 nm, allowing mapping of the local photocurrent. For the detection of the photocurrent, both laser beams are modulated at different frequencies $\Omega_A$ and $\Omega_B$ (both in the kHz range) with a mechanical chopper. The photocurrent is measured either at one of the modulation frequencies or at the sum of both frequencies with a lock-in amplifier.

When detected at one of the modulation frequencies ($\Omega_A$ or $\Omega_B$), the measured photocurrent corresponds to the average photocurrent generated by the modulated pulse train. Using this detection scheme, we obtain the spatially resolved photocurrent image shown in Figure 1(b). The photocurrent is generated at the metal/graphene interfaces due to a potential difference between the graphene covered by the metal contacts and the uncovered part[14-16]. The Fermi level of the detector is fixed and determined by the source and drain contacts. While the carrier density in the covered graphene is given by doping of the metal contacts and is insensitive to external fields, the carrier density in the uncovered part depends on the electric field that is provided by a back gate. Thus, if the gate bias is adjusted such that the doping in the covered part differs from the gate induced doping in the uncovered part, charge redistribution leads to band bending at the metal/graphene interfaces. Hence, a local electric field is established at the interfaces, and under optical illumination, photocurrent is generated without application of a bias between source and drain[14-16]. At monolayer/bilayer[17] and pn-junction[18] graphene interfaces, also a thermal contribution to the photocurrent was reported, that we expect to play a minor role in our devices.

When detected at the sum frequency $\Omega_A + \Omega_B$, the photocurrent corresponds to the photodetector autocorrelation signal containing the intrinsic response time, provided that the laser pulse energy is sufficient to reach the saturation regime of the detector. In graphene, this regime is a result of Pauli blocking and takes place if the photogeneration rate is comparable to the rates of energy relaxation and recombination[19, 4]. The concept of this measurement becomes apparent with the following consideration. If $n_A$ and $n_B$ denote the number of carriers generated by pulse A and B, respectively, and if the reduction in the photogeneration of carriers by pulse A (B) due to saturation (Pauli blocking) caused by pulse B (A) is considered by the factors $(1-c_B)$ and $(1-c_A)$, then the total number of photogenerated carriers can be expressed as $(1-c_B) n_A + (1-c_A) n_B$. If one further assumes that pulse A is modulated with frequency $\Omega_A$ and pulse B with $\Omega_B$, then $c_B n_A$ and $c_A n_B$ are modulated with frequencies $|\Omega_A \pm \Omega_B|$. Thus, when detecting at the sum frequency (or, alternatively, the difference frequency) only the terms $c_B n_A$ and $c_A n_B$ are measured. In the linear regime of the photodetector $c_A$ and $c_B$ are zero since there is no saturation. Consequently the measured signal is zero except if the pulses interfere. In this case the measured signal corresponds simply to the interferometric first order autocorrelation of the ultrashort laser pulses[20]. However, in the nonlinear regime, i.e. when the pulse energy reaches the threshold of saturation[19, 4], $c_A$ and $c_B$ are non-zero, and we are able to extract the response time of the photodetector.

A typical nonlinear autocorrelation signal is shown in Figure 2(a) (logarithmic scale). In addition to the short undersampled interferometric part of the ultrashort laser pulses at zero time delay [labeled (i)], a response tail of the photocurrent is clearly present. The tail is much longer than the laser pulse duration, which was confirmed by pulse duration measurements with a commercial second-harmonic autocorrelator (See Figure 2(b)). It consists of two contributions, one on a sub-picosecond timescale immediately after the temporal overlap of both laser pulses [labeled (ii)], and one on a picosecond timescale [labeled (iii)]. Optical interband excitations with ultrashort pulses result into a non-equilibrium carrier distribution of electrons in the conduction band and holes in the valence band. The excitation is followed by an equilibration process comprising carrier-carrier and carrier-phonon scattering. Although a complete picture of the equilibration dynamics requires further investigations, recent ultrafast pump-probe[21–28] and photoluminescence[29] measurements indicate the following scenario. After photogeneration, the carriers thermalize among themselves on a timescale of tens of femtoseconds via very rapid carrier-carrier scattering leading to separate electron and hole distributions in the conduction and valence bands with nonzero Fermi levels at elevated temperatures. This process is too fast to be resolved in our experiment. Subsequently, a phonon-mediated cooling process of the quasi-equilibrium distributions follows on a 100-femtosecond-time scale. We associate the sub-picosecond contribution (ii) in Figure 2(a) with this cooling process. Eventually, electron-hole recombination establishes a single equilibrium distribution with the Fermi level at the Dirac point in the picosecond regime (See Figure 2(c)). At the same time, (part of) the excess photocarriers get swept out by the electric field built up in the band bending region (See Figure 2(d)). The photocarrier transit time is determined by the mobility of the carriers as well as by the width of the graphene region where photocurrent is generated and the potential difference in this region. Both processes – carrier recombination and transport – are expected to occur on a picosecond timescale in graphene. The total intrinsic response time is given by $t_r^{-1} = t_{rec}^{-1} + t_t^{-1}$, where $t_{rec}$ denotes the recombination time and $t_t$ is the carrier transit time. We attribute the slow component (iii) in Figure 2(a) to the intrinsic response time $t_r$ of the photodetector which describes the material response. Generally, in photodetectors either $t_{rec}$ or $t_t$ must be short for high-speed operation. For instance, in metal-semiconductor-metal photodetectors based on the GaAs material system, high-speed operation can be achieved either by reduction of $t_{rec}$ with use of low-temperature-grown GaAs containing many defects[30], or by reduction of $t_t$ with electrode distances in the sub-100-nm range[31]. In graphene both, $t_{rec}$ and $t_t$, are short. This is why we can expect high-speed operation. In addition to the material response, the circuit response determines the high-speed performance of a photodetector. It is a result of the parasitic capacitance and inductance of the metallization surrounding the photodetection material and depends on the exact geometry of the device. However, since the electrical signals in our experiment are measured at very low frequencies (in the kHz range), the circuit response does not play a role, and we obtain the material's intrinsic response time.

On the logarithmic scale, the bi-exponential character of the signal in Figure 2(a) is apparent. In order to obtain the response time, we perform a linear fit of the rising and falling parts of the slow component (iii) in the measured nonlinear autocorrelation that we expect to be symmetric with respect to zero time delay. We then extract the response time from the average slope of the two fitted curves. We observe a weak asymmetry that we relate to an imperfect overlap of the laser pulses in the experiment, since the generated photocurrent depends strongly on the focus position of the laser beam. Differences between the particular response times due to the asymmetry are, however, small (~0.1 ps). The shortest response time that we extract from the slow component (iii) is $t_r$ = 2.1 ps. This value translates into a bandwidth[32] of $f_c$ = 0.55 / $t_r$ = 262 GHz. The measurement was performed with zero gate voltage and a laser fluence of 12 µJ/cm², sufficient for a nonlinear photoresponse. The position on the sample was chosen according to the maximum current value in the corresponding photocurrent image. The response time $t_r$ is comparable with the ones reported for metal-semiconductor-metal photodetectors based on the GaAs[30,31] and InGaAs[33] material systems. These devices are, however, limited to photon energies above the respective material's bandgap. Graphene-based photodetectors, on the other hand, are expected to operate in a much wider range of photon energies.

In Figure 3(a) a set of autocorrelations is shown that were obtained at different gate voltages (logarithmic scale). In these measurements the sub-picosecond contribution was not fully recorded and is therefore not displayed. For better visibility, a constant offset was introduced between the curves. The corresponding response times $t_r$ are shown in Figure 3(b). A slight shortening of $t_r$ with increasing positive and negative gate voltages is observed. We relate this shortening to a variation of the transit time $t_t$ of the photoexcited carriers. As the electric field strength increases, $t_t$ of the photoexcited carriers across the high-field region decreases. From the rather weak variation of $t_r$ with gate voltage, however, we conclude that the overall response time $t_r = (1/t_t + 1/t_{rec})^{-1}$ is mainly determined by carrier recombination rather than carrier transit. In contrast to the carrier transit time, the recombination time is not expected to vary over the gate voltage range, because a back gate voltage of 10 V corresponds to a shift of the Fermi level of only 90 meV from the charge neutrality point (CNP), i.e. much smaller than the energy of the photogenerated carriers (~0.4 eV – half of the excitation laser energy).

We come to the same conclusion if we compare the measured response time with a simple numerical estimation of the carrier transit time. From the current-voltage characteristics we calculate the mobility of the detector to be $\mu$ ~ 1000 cm²/Vs. With a capacitor model[34] we calculate the potential difference in the graphene region at the graphene/metal interface to be $\Delta V$ ~ 70 mV at zero bias voltage. If we furthermore assume the width of the photocurrent contributing graphene region to be $l$ ~ 200–300 nm[35, 36], we estimate a transit time of $t_t = l^2/(\mu \Delta V)$ ~ 5.7–12.9 ps, i.e. much longer than the measured response time. Taking into account the photodetector's response time $t_r$ = 2.1 ps, we estimate the corresponding range of recombination times to be $t_{rec}$ ~ 2.5-3.3 ps. Consequently, a significant fraction of the photogenerated carriers recombines before the carriers leave the graphene/metal interface region. The internal quantum efficiency of this device is estimated to be $t_r / t_t$ ~ 16–37 %. Similar results have been obtained by Park *et al.*, who estimated that approximately 30 % of the photogenerated carriers contribute to the photocurrent near metal/graphene contacts[16]. In high mobility devices, however, the carrier transit time may become as small, or even smaller, than the recombination time, and the internal quantum efficiency could reach values close to 100 %.

In Figure 4 the gate voltage dependence of the average photocurrent is shown. We obtain a photocurrent gate voltage characteristic that is similar to previous results[10, 14–16]. The photocurrent exhibits two maxima and changes sign at a gate voltage of approximately 5 V. From the current voltage characteristic we determine the Dirac point to be at a gate voltage of about 1 V. Following the procedure in Ref. 15, we estimate a doping of the graphene by the Ti/Au contacts of approximately 50 meV. In addition to the photocurrent, the gate voltage dependence of the measured photocurrent autocorrelation amplitude is shown in Figure 4. This amplitude corresponds to the autocorrelation value with zero time delay between the two pulse trains. The laser fluence, used for illumination in the measurement, was about 115 µJ/cm². Qualitatively, the autocorrelation amplitude follows the photocurrent. For higher bias voltages we observe a deviation from the photocurrent that we attribute partly to charge effects in the $SiO_2$ that were more pronounced when the gate voltage was varied very slowly.

The dependence of response time and autocorrelation amplitude on laser fluence for the MGM detector is shown in Figure 5. In this measurement, the laser fluence was varied from 12 µJ/cm² to 115 µJ/cm². The gate voltage was set to 0 V. The response time of the photodetector increases slightly with increasing laser fluence. Hence, Auger recombination[37] can be excluded as the dominant recombination mechanism. It would show the opposite behavior. Other recombination mechanisms such as plasmon emission[38] or recombination due to intravalley and intervalley optical phonon scattering[39] are more likely. Taking into account an estimated density of photogenerated carriers of about $10^{12}$ cm$^{-2}$ at the lower laser fluence limit and about $10^{13}$ cm$^{-2}$ at the higher laser fluence limit in our experiment, we find qualitative agreement with recent theoretical predictions of the recombination time due to optical phonon scattering[39]. We conclude that response times in the linear regime (at lower laser fluence) should be even shorter than the values obtained in our experiment. Furthermore, we observe the autocorrelation amplitude to correlate with the response time (See Figure 5). The longer the response time, the more carriers accumulate in the high-field region and the stronger the saturation. Since the autocorrelation amplitude depends strongly on the degree of saturation, it follows the response time of the photodetector. This interpretation is also consistent with the variation of response time and autocorrelation amplitude with respect to gate voltage (See Figures 3 and 4(a)). An increasing gate voltage results into a shortening of the response time and accordingly to a reduction of saturation and autocorrelation amplitude.

In summary, we measured the intrinsic response time and the corresponding bandwidth of MGM photodetectors with monolayer graphene, using an ultrafast optical correlation technique. In our experimental approach, we record a nonlinear photocurrent autocorrelation signal that comprises the intrinsic response time of the detector, and we extract the time constants by a fit to the data. Our results indicate that graphene-based optoelectronic devices may have great potential for high-frequency applications in photonics. Moreover, our experimental technique can also be applied to study a variety of other carrier transport phenomena (e.g. carrier transport across pn-junctions or carrier velocity saturation) in graphene on ultrashort time scales.

Acknowledgements

This work was supported by the Austrian Science Fund FWF (SFB-IRON, DK-CoQuS), the Austrian Nano Initiative project (PLATON) and the Austrian Society for Microelectronics (GMe).

36. Carriers are also generated outside this region (~3 μm spot diameter), but those carriers recombine and do not contribute to the photocurrent.
37. Rana, F.; *Phys. Rev. B*, 76, 155431, **2007**
38. Rana, F.; Strait, J. H.; Wang, H.; Manolatou, C.; *arXiv*:1009.2626v1 [cond-mat.mes-hall], **2010**
39. Rana, F.; George, P. A.; Strait, J. H.; Dawlaty, J.; Shivaraman, S.; Chandrashekhar, M.; Spencer, M. G.; *Phys. Rev. B*, 79, 115447, **2009**

Figures

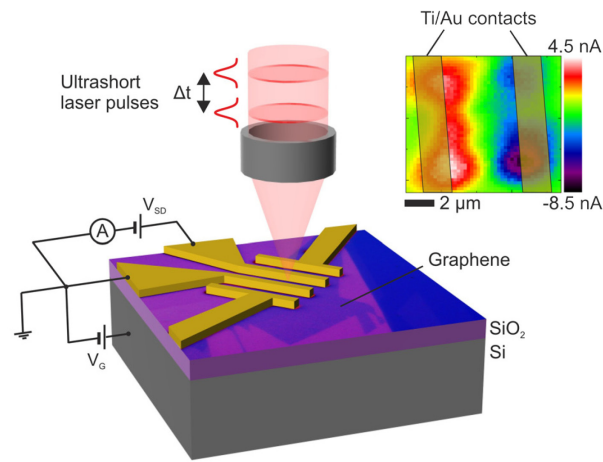

Figure 1. (a) Schematic illustration of the experimental setup and sample structure consisting of three devices. The used laser system has a wavelength of 1.55 μm and a pulse length of 100 fs. (b) Spatially resolved photocurrent image obtained at a gate voltage of 0 V and a laser fluence of 115 μJ/cm².

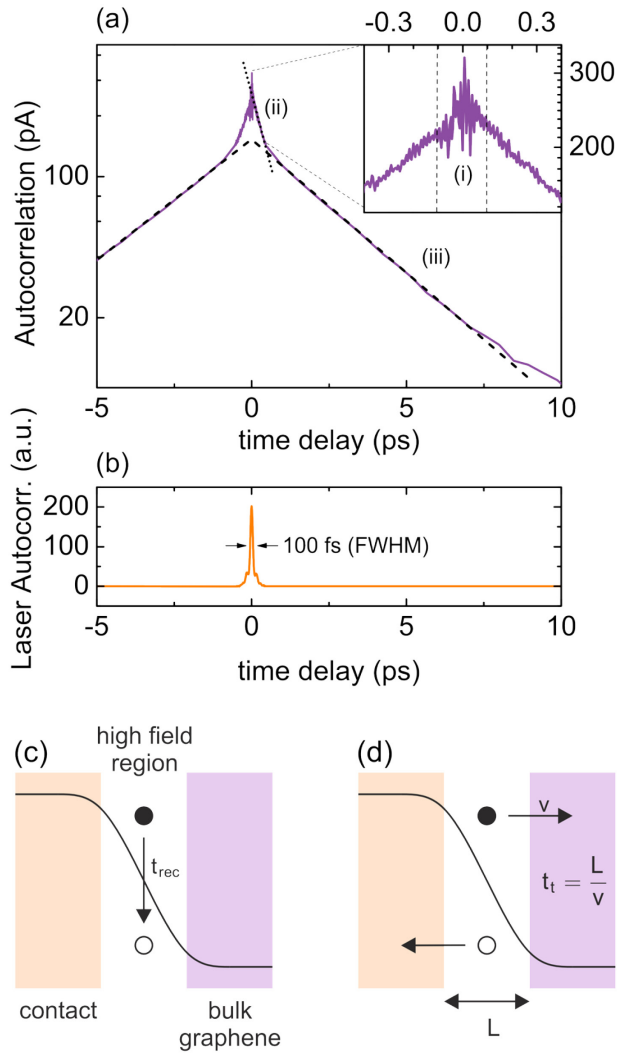

Figure 2. Nonlinear photoresponse to two subsequent ultrashort laser pulses. (a) Photocurrent autocorrelation signal. Note the bi-exponential decay on a logarithmic vertical scale (dashed and dotted lines). Part (i) is due to interference of the lasers pulses. Part (ii) corresponds to a sub-picosecond contribution associated with carrier relaxation via phonons, part (iii) to a contribution on a picosecond timescale connected to the response time of the photodetector. The response time is determined from (iii) by a linear fitting procedure of the right- and left-sided parts (dashed lines) of the autocorrelation function and subsequent averaging. (b) Second order autocorrelation of the laser pulses. (c, d) Processes contributing to the response time: carrier recombination (c) and carrier transport (d).

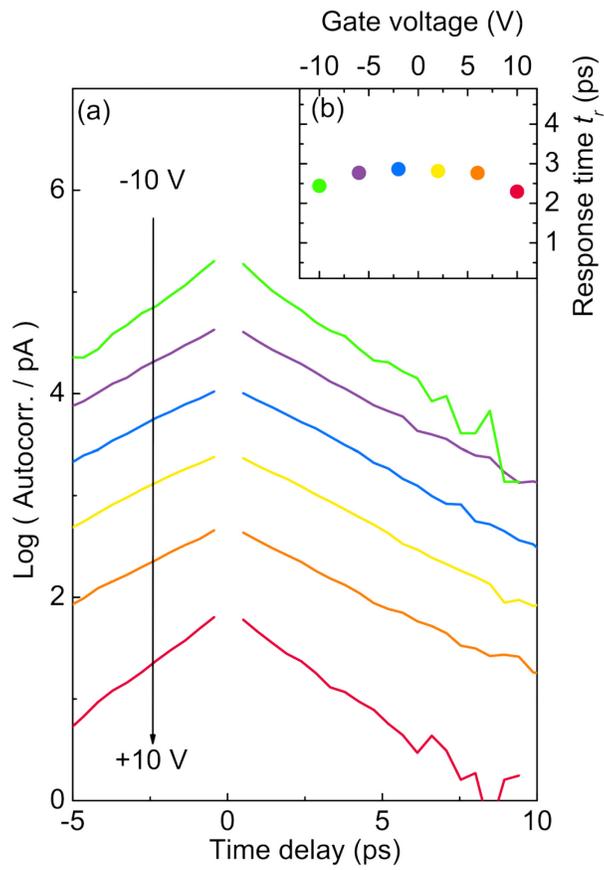

Figure 3. (a) Photocurrent autocorrelations measured at gate voltages from -10 V to 10 V. A constant offset is introduced between the curves for a better comparison. (b) Corresponding response times $t_r$ extracted from the slopes of the photocurrent autocorrelations.

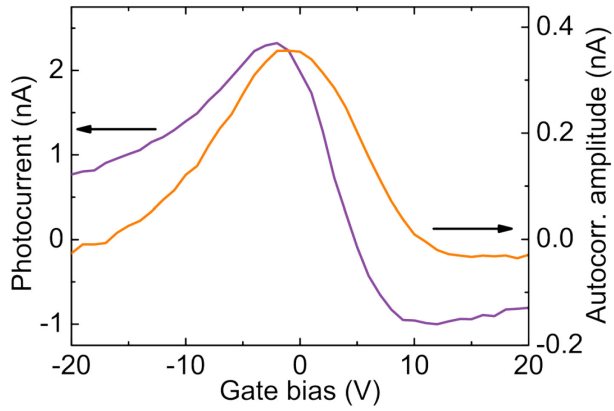

Figure 4. Variation of average photocurrent and photocurrent autocorrelation amplitude with gate voltage. The autocorrelation amplitude is the autocorrelation value at zero time delay between two subsequent laser pulses.

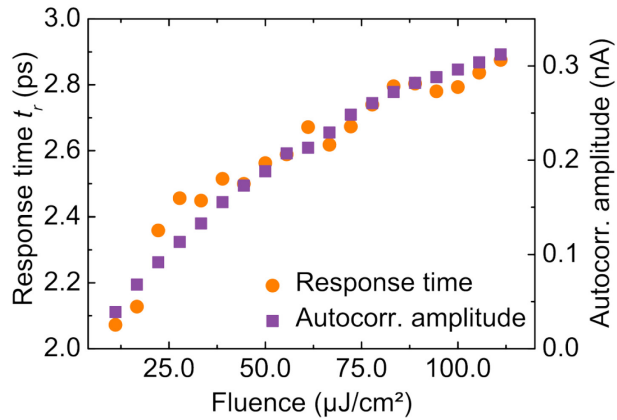

Figure 5. Variation of response time and photocurrent autocorrelation amplitude with laser fluence. The results were obtained at a gate voltage of 0 V.